\documentclass[aps, prd, a4paper, twocolumn,  showpacs, superscriptaddress, nofootinbib, reprint, longbibliography]{revtex4-2}

\usepackage{bm, amssymb, amsmath, mathrsfs,  amsfonts, multirow, float, siunitx, url, lipsum}

\usepackage{booktabs}
\usepackage[usenames]{color}
\usepackage{graphicx, capt-of}

\usepackage{hyperref}
\hypersetup{
	colorlinks=true, 	
	linkcolor=cyan,		
    citecolor=blue,		
}

\newcommand{\nth}{\textsuperscript{th}} 
\newcommand{\norm}[1]{\left\lVert#1\right\rVert}
\newcommand\abs[1]{\ensuremath{\lvert#1\rvert}}
\newcommand\logtwo{\ensuremath{\log_{2}}}

\newcommand{\lnb}[1]{\ln(#1)}

\newcommand\inp[2]{\langle #1 \,|\, #2 \rangle}

\def \match {\mathscr{M} }
\def \Hz	{\mathrm{Hz}}

\def \msun	{{M}_\odot}

\def \imrppv {\mathrm{IMRPhenomPv2}}
\def \imrxp {\mathrm{IMRPhenomXP}}
\def \seobnr {\mathrm{SEOBNRv4}}
\def \aligo	{\mathrm{aLIGO}} 
\def \flow	{f_{\mathrm{low}}}

\def \fhigh	{f_{\mathrm{high}}}
\def \hinj { h_{\mathrm{inj}} }
\def \hrec { h_{\mathrm{rec}} }
\def \hlif { h_{\mathrm{bif}} }

\def \lalinf {\textsc{LALInference}}
\def \bilby {\textsc{Bilby}}
\def \bayeswave {\textsc{BayesWave}}
\def \bayesline {\textsc{BayesLine}}
\def \pesummary {\textsc{PESummary}}
\def \cwb {\textsc{cWB}}

\begin{document}

\title{Nonorthogonal wavelet transformation for reconstructing gravitational wave signals}

\author{Soumen Roy}
\email{soumen.roy@nikhef.nl}
\email{roy.soumen61@gmail.com}
\affiliation{Nikhef, Science Park 105, 1098 XG Amsterdam, The Netherlands}
\affiliation{Institute for Gravitational and Subatomic Physics (GRASP), Utrecht University, Princetonplein 1, 3584 CC Utrecht, The Netherlands}

\date{\today}

\begin{abstract}

Detections of gravitational-wave signals from compact binary coalescences have enabled us to study extreme astrophysical phenomena and explore fundamental physics. A crucial requisite for these studies is to have accurate signal models with characteristic morphologies, which have been challenging for many decades and researchers are still endeavoring to incorporate important physics. Therefore, morphology-independent methods have been developed for identifying a signal and its reconstruction. The reconstructed signal allows us to test the agreement between the observed signal and the waveform posterior samples from parameter estimation. These methods model observed signals using a nearly orthogonal wavelet basis in the frame of continuous wavelet transformation. Here, we propose log-uniform scales to construct the wavelets, which are are highly redundant (non-orthogonal) compared to the conventional octave scales but more efficient for reconstructing the signals at high frequencies. And we introduce a semi-model-dependent reconstruction method using the posterior samples of the events, where we model the signal using Gabor-Morlet wavelets with log-uniform scales. We demonstrate the ability to detect deviation using a numerical simulation of an eccentric binary black hole merger, where the signal in the data does not belong to the search template waveform manifold. Finally, we apply this method to each binary black hole merger event in GWTC-1. We have found that the signal produced by the GW150914 event has 96\% agreement with the waveform posterior samples. As the detector sensitivity improves and the detected population of black hole mergers grows, we expect the proposed method will provide even stronger tests.

\end{abstract}

\maketitle

\section{Introduction}

Detecting gravitational waves (GWs) from the merging compact binaries consists of black holes and neutron stars has ushered in a new era of observational astronomy~\cite{gw150914, gw170817}. So far, the network of Advanced LIGO and Advanced Virgo Detectors has observed more than 90 of these events~\cite{ GWTC1, GWTC2, 2021arXiv211103606T, Venumadhav:2019tad, Nitz:2018imz, Nitz:2020oeq}. Observations of compact binary mergers allow us to design a range of novel tests of general relativity (GR) in the strong-field regime~\cite{TGR-GW150914, TGR-GWTC1, TGR-GWTC2}. A current limitation on tests of beyond-GR physics with compact binary coalescences is the lack of understanding of the strong-field merger regime in nearly all modified theories of gravity. Thereby, we generally perform the tests assuming GR to be the null hypothesis or evaluate the consistency of the data with predictions from the theory. At the same time, the latter test can reveal if any unknown binary parameter influences the observed signal.

The continuous wavelet transformation (CWT) based time-frequency analysis has played a crucial role in identifying the gravitational wave events and detecting the deviation from general relativity. The most potential gravitational wave sources, such as compact binary mergers, supernovae explosions, produce non-stationary signals. The CWT is a powerful analysis tool that allows us to obtain a time-frequency localized projection of a non-stationary signal. This procedure calculates the wavelet coefficients by performing a scalar product between the signal and wavelet basis. These coefficients are used not only for time-frequency analysis but also to reconstruct the original time-domain signal. The existing time-frequency-based methods such as coherent wave burst ($\cwb$)~\cite{Klimenko_2004, Klimenko:2005xv, Klimenko_2008}, $\bayeswave$~\cite{Cornish_2015, Cornish:2020dwh}, and X-pipeline~\cite{Sutton:2009gi, Chatterji:2006nh} are designed for searching unmodeled gravitational-wave bursts and reconstructing the signal from the coherent response of a network of detectors. 
The previous studies reported satisfactory agreement between the reconstructed signal and the estimated waveform of binary black hole mergers for the louder events~\cite{GWTC1, Ghonge:2020suv, Salemi:2019uea}.


The conventional burst search method identifies a cluster of pixels from the time-frequency map of a detector's strain data if each of those pixels has more power than one expects from the noise alone, called triggers. If the triggers from a network of detectors are coincident, then the method claims a detection of a signal~\cite{Anderson:2000yy}. The $\cwb$ method first constructs a multi-resolution time-frequency map of the strain data using Wilson-Daubechies-Meyer wavelets~\cite{Necula:2012zz} and uses the same strategy to detect the events. After that, the method employs a coherent framework of the constrained maximum likelihood approach to yield the event properties: sky location, wave polarization, and signal reconstruction~\cite{Klimenko:2005xv, Klimenko:2015ypf}. Whereas the $\bayeswave$ method employs a framework of Bayesian statistics without relying on any prior assumption of waveform morphology~\cite{Cornish_2015}. This method reconstruct the gravitational wave signals and instrumental noise, where both of them are characterized as a superposition of Gabor-Morlet wavelets or chrirplets~\cite{Millhouse_2018}. The number of time-localized wavelets and their parameters are determined via Reversible Jump Markov Chain Monte Carlo algorithm. After that, whether the likelihood of the event is favoured to a true gravitational wave signal or instrumental glitch is determined by using Bayesian model selection strategy. Finally, the posterior samples are used to reconstruct the signal as a superposition of the wavelets.

A wavelet can be regarded as a time-frequency localized bandpass filter. The bandwidth of which is determined by the properties of that wavelet. The reconstructed signal has an inevitable noise contribution that passes through those wavelet filters. The match between the reconstructed signal and the estimated waveform increase with the signal-to-noise ratio (SNR). However, we would never achieve absolute agreement (unit match) even if our predicted theory is complete because of the additional noise that passes through the coherent wavelets.


 In this paper, we propose a quasi model-dependent method for reconstruction of signal from noisy data. We identify the essential wavelets using the binary black holes merger waveforms estimated using the standard Bayesian analysis~\cite{Rover:2006ni, vanderSluys:2008qx, Veitch:2014wba, Skilling:2006gxv}. The unmodelled reconstruction methods identify the wavelets that are coherent across the network of detectors. The collection of those coherent wavelets in the time-frequency plane form a cluster. Similarly, the essential wavelets form a cluster in the time-frequency plane which can be collectively looked as a single patch with an irregular boundary. 


The patch area determines how much noise persists in the reconstructed signal. The wavelets are discretely placed over the time-scale plane, where the patch area is determined by the placement method and sparseness. The dyadic grid placement is considered to be the most efficient method, leading to the construction of an orthonormal wavelet basis. This placement assumes an output of octave decomposition (power-of-two logarithmic) to construct the grid. The sparseness of the grid imposes a limit on the acceptable loss in signal characteristics. We target to achieve maximum sparseness and still have an adequate representation. A highly sparse grid might provide a sufficient signal representation, but the area covered by the essential wavelets would be larger than the case of a denser grid. On the contrary, an over-dense grid can not reduce the patch area beyond a specific limit and also increases the computational cost.

In this paper, we propose a log-uniform scale to place the wavelets over the time-scale domain. 
Further, we derive an inverse wavelet transformation formula for log-uniform scale. The wavelets with such scales are highly redundant (i.e., over complete), but they can further reduce the area covered by the essential wavelets. As a result, We can reduce the noise contribution in the reconstructed signal, which leads to a more accurate signal reconstruction.  Moreover, the essential wavelets with log-uniform scale can provide a more appropriate signal representation than the octave scale for the high-frequency linear chirp signals. 
It implies that the log-uniform scale is more efficient for reconstructing the signals from low mass binary systems.


This paper is organized as follows: Section~\ref{sec:WaveletReconstruction} describes the reconstruction methods using the continuous wavelet transformation with octave and log-uniform scales; 
Section~\ref{sec:performance} demonstrates the performance of the reconstruction methods for the linear chirp signals and gravitational wave signals in simulated Gaussian noise; Section~\ref{sec:Deviation} describes the efficiency of the wavelet reconstruction method when a signal in the data does not belong to the manifold of search template waveform by injecting an eccentric waveform; Section~\ref{sec:GWTC1Ananlysis} exhibits results of our analysis for binary black hole merger events observed by the Advanced LIGO and Virgo during the first and second observation runs. Finally, we conclude in Section~\ref{sec:Consclusion}.


\section{Reconstruction formulae for non-orthogonal wavelet}
\label{sec:WaveletReconstruction}
The CWT of a continuous signal $x(t)$ is a linear mapping onto a time-scale space of wavelets:
\begin{equation}
\label{eq:cwt}
\begin{split}
    X( a, b) & = \int x(t) \, \Psi^\ast_{a, b}(t) \, dt \\
    &=\frac{1}{\sqrt{a}} \int x(t) \, \psi^\ast\left(\frac{t-b}{a}\right) \: dt ,
\end{split}
\end{equation}
where, $\psi^\ast$ is the complex conjugate of the shifted and scaled  version of the time-localized mother wavelet $\psi$, in which  $a$ and $b$ are the scale and time-shift parameters, respectively. Therefore, the CWT provides a time-scale representation of the signal by performing a sliding cross-correlation with a continuous family of wavelets. The quantity $1/\sqrt{a}$ outside the integral is an energy normalized factor. It assures that each wavelet has the same energy, whatever the value of scaling and shift.  The function $\psi(t)$ must satisfy a set of mathematical criteria to be a wavelet, it must have finite energy, 
\begin{equation}
    E_\psi = \int_\infty^\infty \abs{ \psi(t) }^2 \: dt < \infty, 
\end{equation}
and must hold the admissibility condition,
\begin{equation}
\label{eq:admissibility_constant}
    C_{\psi} = \int_\infty^\infty \frac{ \abs{\tilde{\psi}(\omega) }^2 }{\omega} \: d\omega < \infty ,
\end{equation}
where, $\tilde{\psi}(\omega)$ is the Fourier transform of $\psi(t)$, $\omega = 2\pi f$ is the angular frequency, and $C_{\psi}$ is called the admissibility constant. The above condition implies that $\tilde{\psi}(\omega)$ is endeavoured to zero faster than $\omega$ and must not have zero frequency component, $\tilde{\psi}(0)=0$. If the wavelet function satisfy both the criteria, then the inverse wavelet transform can be described as a superposition of the dual wavelets:
\begin{equation}
\label{eq:icwt1}
    x(t) = \frac{1}{C_\psi} \int_0^\infty \int_{-\infty}^{\infty} X(a, b) \: \Psi_{a, b}(t) \: \frac{1}{a^2} \, db \, da
\end{equation}
An alternative approach of inversion formula was found by Morlet, we can even choose a a completely different wavelet function, Dirac  $\delta$-function  $\delta\left((b-t)/a\right)$  instead of the analysing wavelet, and that leads to a single integral inverse formula~\cite{Farge-1992},
\begin{equation}
\label{eq:icwt2}
    x(t) =  \frac{2}{C_\delta} \int_0^\infty \Re\left\{ X(a, t)\right\} \frac{1}{a^{3/2}}  \, da , 
\end{equation}
where $C_\delta$ is the admissibility constant of the $\delta$-function that can be computed using the Eq.\eqref{eq:admissibility_constant}.
In this work, we use this single integral inversion formula for reconstructing the gravitational wave signals.

The wavelet function can be regarded as a impulse response of a bandpass filter. The associated frequency of the wavelet can be treated as frequency value in the time-frequency domain, which is known as pseudo-frequency ($f_p$). It depends on the central frequency ($f_c$) and the scale parameter of the wavelet, 
\begin{equation}
    f_p = f_c/a
\end{equation}
This equation can used to represent the CWT in time-frequency frame as like the short-time Fourier transform.

One of the most commonly used mother wavelet is the Morlet wavelet~\cite{Morlet-1984}, which is a complex non-orthogonal wavelet. As the complex wavelet can separate the phase and amplitude components associated with the signal, it is more suited to determine the instantaneous frequency. The Morlet wavelet consists of an harmonic oscillation with Gaussian window,
\begin{equation}
    \psi(t) =  A_{f_c} \pi^{-1/4} \left( e^{i 2\pi f_c t} - e^{-(2\pi f_c)^2/2} \right) e^{-t^2/2},
\end{equation}
where $A_{f_c} = (1 - e^{-4\pi^2 f_c^2} - 2e^{-3\pi^2 f_c^2})^{-1/2}$ the normalization constant. The second term in the bracket is a correction to preserve the admissibility condition. It ensures that the zero-frequency component vanishes. In practice, we ignore this term as it is approximately zero for the values of $f_c \gg 0$. Note that the above equation does not contain the time-shift and scale parameters. When we analyze a signal, we replace the variable $t$ with $(t-b)/a$. 


The central frequency $f_c$  is a crucial parameter in the time-frequency analysis to regulate the tradeoff between temporal precession and spectral resolution. 
It controls the number-of-cycles of the wavelet without modifying the shape of Gaussian window. A large value of central frequency leads to an increased temporal precision at the cost of decreased spectral precision and vice versa for a smaller value of central frequency. It is impossible to achieve simultaneously good precision in time as well as in frequency. A rule of thumb can be proposed for analyzing the gravitational waves from compact binary mergers.
The signals from low-mass systems spend several tens of seconds or more than a hundred seconds in the bandwidth of Advanced LIGO-like detectors, which look like a long-duration chirp signal. A larger value of central frequency is convenient to achieve an overall good time-frequency precision for those systems. On the other hand, a smaller value of central frequency is suitable for high mass systems as their signals are short-duration bursts with a tiny chirp.

\subsection{Choice of wavelet scales: octave}
The CWT generally suffers due to the redundancy at large scales, where the neighboring wavelets are highly correlated. Once a wavelet function is chosen, it is important to choose a tightest set of scales that forms an orthonormal wavelet set. The key mathematical criteria to choose a set of discrete wavelets is that every function $f\in L^2(\mathbb{R})$\footnote{$L^2(\mathbb{R})$ denotes the space of square integrable functions on the real line ($\mathbb{R}$). } must be fully expressed as a superposition of those wavelets i.e., the set is complete in $L^2(\mathbb{R})$. The conventional approach of choosing the scales is an output of octave decomposition~\cite{Grossmann-1989, APracticalGuidetoWaveletAnalysis},
\begin{equation}
\begin{split}
        a_j & = a_0 \: 2^{j \delta j}, \ \ j = 0, 1, 2,..., J \\
        J &= \delta j^{-1} \logtwo{\left(N \delta t/ a_0\right)}, 
\end{split}
\end{equation}
where $N$ is the total number of samples in $x(t)$, $\delta t$ is sampling interval, $a_0=2\delta t \, f_c$ is the smallest resolvable scale, and $\delta j$ is the spacing between the discrete scales. For Morlet wavelet, an adequate scale resolution can be achieved for the values of $\delta j \lessapprox 0.5$. The quantity $J$ that determines the highest value of the scale is associated with the Nyquist frequency. The octave scale leads to a simpler implementation of inverse discrete wavelet transformation as the quantity $da/a = a_0 \delta j \lnb{2} $ is a constant. Now, the inverse formula Eq.~\eqref{eq:icwt2} yields, 
\begin{equation}
\label{eq:ReconstructionOctave}
    x_n = \frac{\delta j \delta t^{1/2}}{C_{\delta} \psi_0(0)} \sum_{j=1}^{J} \frac{\Re{ \left\{ X_n (a_j) \right\} }}{ a_j^{1/2}}
\end{equation}

\subsection{Choice of log-uniform scale}

For non-orthogonal wavelet analysis, one can adopt an arbitrary set of scales to build up a more complete picture if that set satisfy the completeness criteria. We propose a log-uniform scale,
\begin{equation}
\label{eq:log-uniformSacle}
\begin{split}
    \frac{1}{a_j} &= \frac{1}{a_{j-1}} - \delta\ell \\
                 &= \frac{1}{a_0} - j\delta\ell .
\end{split}
\end{equation}
The quantity $\delta\ell=\delta f_p/f_c$ stands for the inverse of spacing between two discrete wavelets and smallest resolvable scale $a_0 = f_c/\delta f_p$. In this work, we choose $\delta f_p = 1/T$, where $T$ is duration of the time-series. 
The log-uniform scale leads to an uniformly spaced pseudo-frequency. In this formalism, Morlet wavelet transformation is equivalent to the scaled Gabor transformation as the pseudo-frequencies are uniformly spaced. For log-uniform scale $da/a^2 = \delta\ell$, thus, the discrete inverse formula Eq.~\eqref{eq:icwt2} yields,
\begin{equation}
\label{eq:Reconstructionlog-uniform}
x_n = \frac{\delta\ell \, \delta t^{1/2}}{C_{\delta} \, \psi_0(0)} \sum_{j=1}^{J}  a_j^{1/2} \: \Re{ \left\{ X_n (a_j) \right\}}    
\end{equation}

\subsection{Identifying the essential wavelets}

The conventional method for identifying essential wavelets is a straightforward approach, where one discards the low-magnitude wavelet coefficients by applying a threshold, which is known as wavelet thresholding.~\cite{Donoho-1994, Johnstone-1997}. This approach aims to remove the noise from data without affecting the basic features of the signal. In this work, we propose a slightly different approach. We employ the modelled waveforms that are obtained from the standard Bayesian parameter estimation analysis. We compute wavelet coefficients for a waveform and set a threshold on coefficient value to select the essential wavelets. The threshold is chosen such that the resultant power from the essential wavelets is equal to a percentage of total power of the spectrogram. We calculate the threshold ($E_{\ast}$) for a given value of fractional power loss ($ R_{\ast}$) by solving an equation,
\begin{equation}
\label{eq:ImportantWavelets}
R_{\ast} = 1 -  \sum_{n, j} \abs{X(n, j)}^2 \, \left[ \abs{X(n, j)}^2 \geq E_{\ast} \right] \bigg/ \sum_{n, j} \abs{X(n, j)}^2.
\end{equation}  
We shall call that $R_\ast$ is the spectral loss parameter. As the number of essential wavelets is considerably fewer than the total number of wavelets, a significant amount of noise can be removed while preserving the basic features of the signal.


\begin{figure}
\centering
  \includegraphics[width=0.48\textwidth]{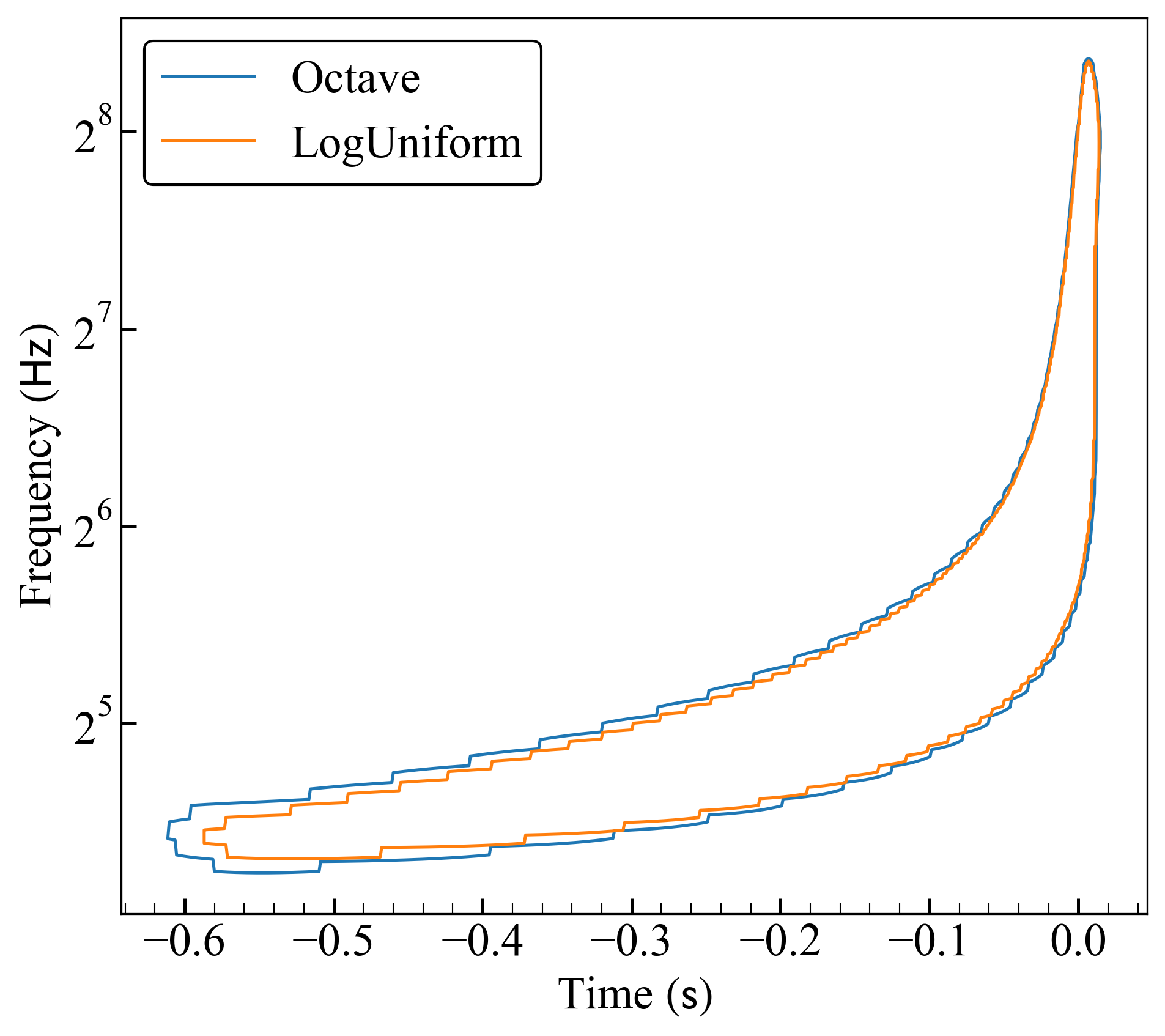}
  \caption{ An illustration of the area covered by the cluster of essential wavelets is produced using a gravitational waveform of a nonspinning equal mass binary system with a chirp mass of $30 \msun$. The areas bounded by the log-uniform wavelets and octave scale wavelets  are 13 and 15, respectively.  For this computation, we have assumed $\aligo$~\cite{aLIGO_ZDHP} noise curve with a fixed lower cutoff frequency of $20\Hz$, the waveform is generated using $\seobnr$ model~\cite{Bohe:2016gbl}. We have considered $R_\ast=0.05$ for determining the essential wavelets.}
  \label{fig:area_comparison}
\end{figure}

\subsection{Occupied area by the essential wavelets}
Reconstructing a signal using the essential wavelets trades with the sensitivity, by which we are able to remove the noise from data. However, the reconstructed signal would always contain a nominal amount of inseparable noise, which passes through the wavelet filters. As a wavelet is regarded as a time-localized bandpass filter, it can be seen as a patch on time-frequency plane. The area (time-frequency bandwidth) of that patch can be used to determine the amount noise released through that wavelet. Therefore, the total amount of inseparable noise can be estimated by calculating the total area covered by the essential wavelets.

The area on the time-frequency plane covered by a cluster of wavelets $\{ (b_i, a_i) \}_{i=1}^{n_0}$ is related to their placement. For octave scale wavelet with a central frequency of $f_c$, the covered area is:
\begin{equation}
\begin{split}
A_{\rm{Octave}} & = \delta t  \:  a_0 \delta j\lnb{2} \sum_{i=1}^{n_0}   a_i \\
& = 2\lnb{2}  \: \delta j \left( f_c \delta t\right)^2 \sum_{i=1}^{n_0}  1 /  f_{p i}, 
\end{split}
\end{equation}
where, $f_{pi}$ is the $i\nth$ pseudo frequency, $f_{p i}=f_c/a_i$. Whereas, the time-frequency area governed by a cluster of wavelets $\{ (b_i, a_i) \}_{i=1}^{n_0}$ with log-uniform scale is:
\begin{equation}
A_{\rm{log-uniform}} =   \sum_{i=1}^{n_0}  \delta t /T .
\end{equation}

 Fig.~\ref{fig:area_comparison} shows that log-uniform scale occupy a smaller area than the octave scale.  The figure indicates that the log-uniform scale allows $\sim 13\%$ less noise in the reconstructed signal than the octave scale for a high mass system, can provide a nearly identical signal representation.

\section{Performance of the reconstruction procedures}
\label{sec:performance}
In this section, we demonstrate the performance of the wavelet based signal reconstruction and compared between the choice of Octave and log-uniform scale. First, we exhibit that the essential wavelets are adequate to represent a chirp signal containing a broad range of frequency. Second, we carry out injection analysis to evaluate the performance for real data analysis. The gravitational wave signals are drawn from the binary black merger and added to a simulated Gaussian noise.

%
\begin{figure}
\centering
  \includegraphics[width=0.48\textwidth]{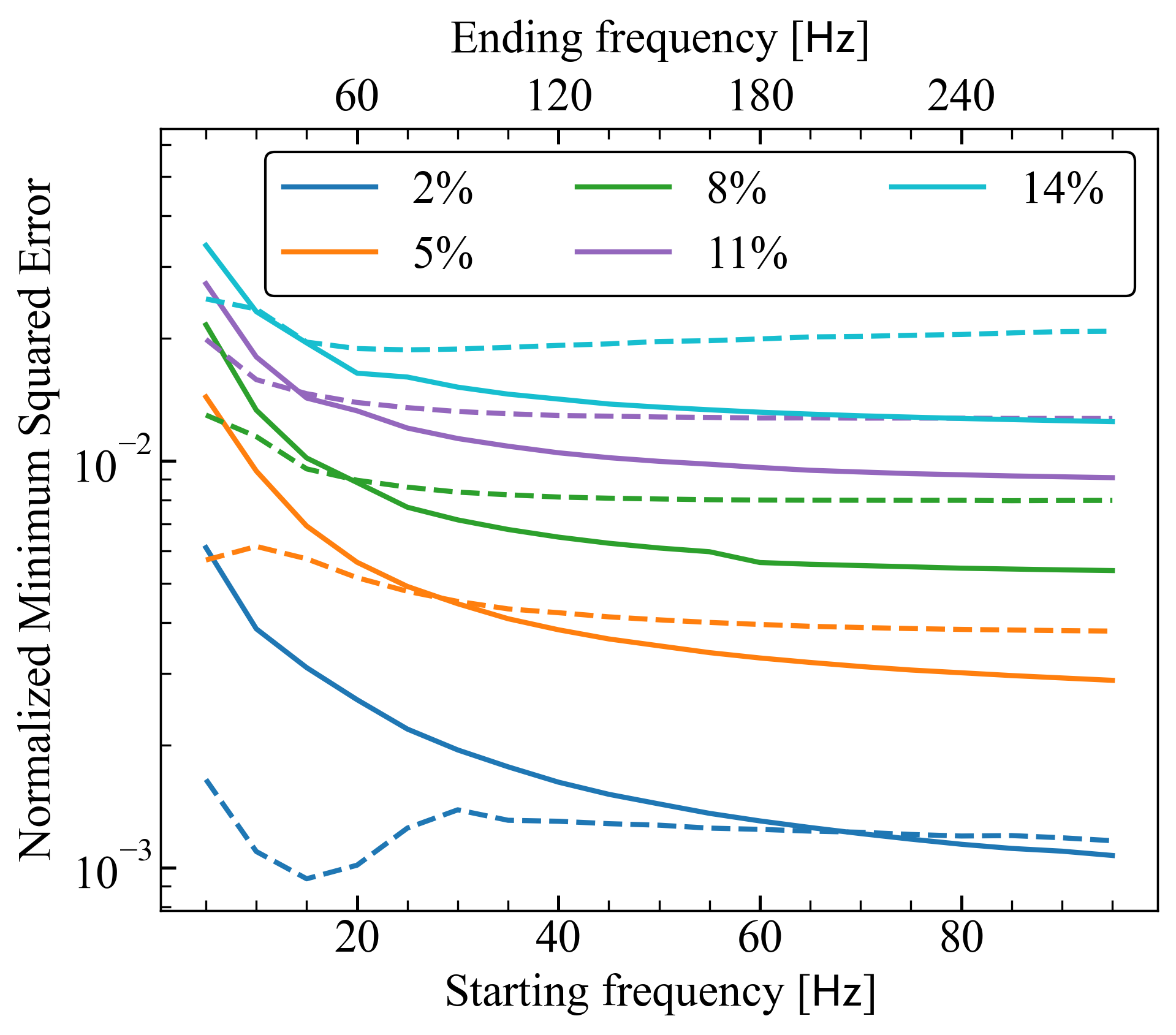}
  \caption{ Performance of octave and log-uniform scale wavelets reconstruction for the linear chirp signals. The signals are generated using Eq.~\eqref{eq:simple_chirp}.  The y-axis represents the normalized minimum squared error in the reconstructed signal. The solid line and dashed line correspond to the log-uniform scale and octave scale, respectively. The bottom and top x-axes represent the starting and ending frequency of the chirp, respectively. The legend represents a set of spectral loss parameters ($R_{\ast}$) as described in Eq.~\eqref{eq:ImportantWavelets}. We used those values to choose the essential wavelets. }
  \label{fig:power_comparison}
\end{figure}

\subsection{Reconstruction of chirp signal without noise }
\label{sec:RecChirpSignal}
We consider a simple chirp signal with a constant amplitude and phase is upto a quadratic order in time $(t)$,
\begin{equation}
\label{eq:simple_chirp}
x(t) = \sin\left(  2\pi f_0t + 2\pi f_0t^2 \right),
\end{equation}
where $f_0$ is the starting frequency of the signal 
and the time range is chosen to be between 0 and $2\, \si{\second}$. Therefore, the ending frequency of the chirp is $3f_0$. This type of chirp signal is adequate to exhibit the performance of the reconstruction method over a broad range of frequencies.  We consider five different cases to identify the essential wavelets. 
For each case, the total power contained in the essential wavelets equals a fraction of signal power. 

It is inevitable to have an overall amplitude loss since the number of essential wavelets is a small subset of the set of wavelets representing the whole time-scale space. 
However, for a given value of $R_{\ast}$, we can set the overall amplitude by looking at the amplitude loss when determining the essential wavelets. An alternative approach is to define normalized minimum squared error (NMSE) for a reconstructed signal $x_{\rm{rec}}(t)$,
\begin{equation}
\label{eq:NMSE}
 \mathcal{E}_{\rm{NMSE}} =  \int \Big| x(t)/\norm{x} - x_{\rm{rec} }(t)/\norm{x_{\rm{rec}}} \Big|^2 \: dt   ,
\end{equation}
where, the symbol $\norm{\cdot}$ denotes the norm. We use this above equation to quantify the adequateness of the essential wavelets to characterize a linear chirp signal.

 In Fig.~\ref{fig:power_comparison}, we illustrate the accuracy of the wavelet reconstruction of the linear chirp signals. The solid and dashed lines correspond to the log-uniform scale and octave scale, respectively.  We consider five different cases of fractional loss in spectrogram power to select the essential wavelets used for reconstruction. The reconstruction of high-frequency chirps using log-uniform scale wavelet is more accurate than the octave scale and vice versa for the low-frequency chirps.

\subsection{Reconstruction of gravitational wave signal in simulated noise }
\label{subsec:gwsimnoise}
We estimate the performance of the reconstruction methods for gravitational wave signals from compact binary mergers of equal mass nonspinning black holes. For each case, an identical signal is injected in many noise realizations of stationary Gaussian distribution weighted by Advanced LIGO zero-detuned high-power ($\aligo$) design sensitivity~\cite{aLIGO_ZDHP}. The waveforms are generated for an wide range of chirp mass between $10 \msun$ and $40 \msun$ using $\seobnr$ model~\cite{Bohe:2016gbl} with a fixed lower cutoff frequency of $20\Hz$.  To determine the reconstruction accuracy with signal-to-noise ratio (SNR) of the injection (injected SNR), we choose a set of values between 5 and 50. The injected SNR ($\rho_{\rm{inj}} $) of a waveform $(h_{\rm{inj}})$ is defined as:
\begin{equation}
\rho_{\mathrm{inj}}^2 = 4  \int_{\flow}^{\fhigh}  \frac{ \tilde{h}^{\ast}_{\mathrm{inj}}(f) \: \tilde{h}_{\mathrm{inj}}(f) }{S_n(f)} \, df , 
\end{equation}
where, $\tilde{h}_{\mathrm{inj}}(f)$ denotes the Fourier transform of $\hinj(t) $ and $S_{n} (f)$ denotes the one-sided detector noise power spectral density.

\begin{figure}
\centering
  \includegraphics[width=0.48\textwidth]{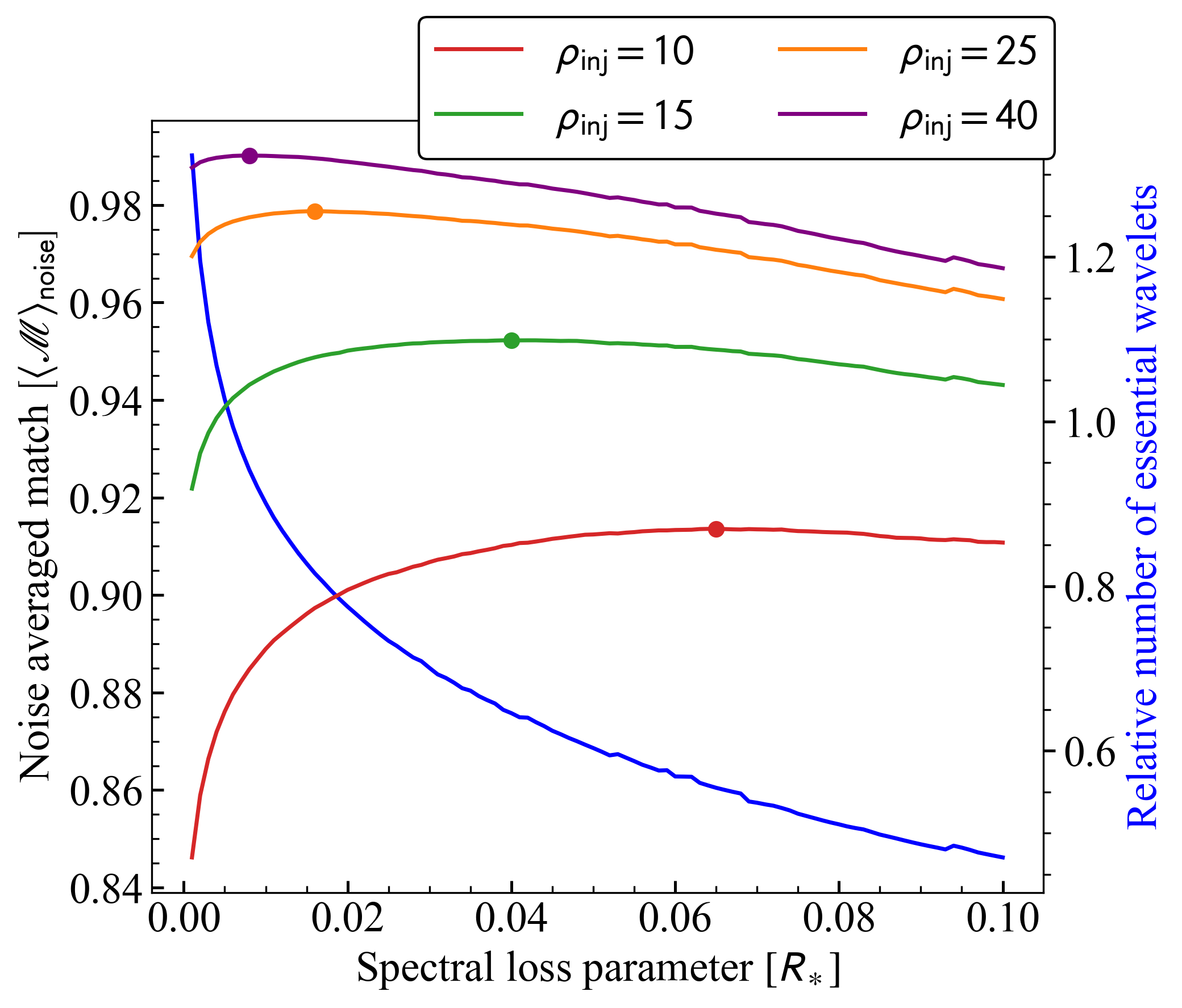}
  \caption{ The noise averaged match ( $ \langle \match \rangle_{\rm{noise}} $ ) as a function of spectral loss parameter  ($R_\ast$) is produced using a gravitational waveform of a nonspinning equal mass binary system with a chirp mass of $30 \msun$. The solid dot on the curve indicates its maxima. The right y-axis shows the relative number of essential wavelets  ($\bar{N}_W$) as a function of $R_\ast$. The unit value of  $\bar{N}_W$ corresponds to a 99.9\% match between the original and reconstructed signal for the zero noise case. These results are produced assuming the $\aligo$ noise curve with a fixed lower cutoff frequency of $20\Hz$. The waveforms are generated using the $\seobnr$ model.}
  \label{fig:spec_loss}
\end{figure}

We reconstruct the signal from noisy data using the log-uniform scale wavelet transform, where the essential wavelets are selected using the Eq.~\eqref{eq:ImportantWavelets}. 
To quantify the signal reconstruction accuracy, we calculate the \textit{match} between the injected signal $\hinj$ and reconstructed signal $\hrec$, which is defined as a inner product between two normalized waveforms ($\hat{h}_\ast = {h_\ast}/\sqrt{ \inp{ h_\ast }{ h_\ast } }$): 
\begin{equation}
\label{eq:overlap}
\match (\hinj, \hrec) := \inp{ \hat h_{ \mathrm{inj} } }{ \hat h_{ \mathrm{rec} } } , 
\end{equation}
where the term with angular brackets represents the following inner product,
\begin{equation}
 \inp{ h_{ \mathrm{inj} } }{ h_{ \mathrm{rec} } } = 4 \Re \int_{\flow}^{\fhigh}  df \, \frac{ \tilde{h}^{\ast}_{ \mathrm{inj} }(f) \, \tilde{h}_{ \mathrm{rec} }(f) }{S_n (f)} , 
\end{equation}
where $ \tilde{h}_{ \mathrm{inj} }(f) $ and $\tilde{h}_{ \mathrm{rec} }(f) $ denote the Fourier transform of $\hinj(t)$ and $\hrec(t)$, respectively.

The match between two waveforms varies between -1 and 1, depending on their correlation but not their overall amplitudes. A match value of 1 indicates a perfect positive correlation such that two waveforms change with equal proportion, and 0 means no correlation. The equal proportion changes with reverse direction indicate perfect negative correlation, for which match value is -1. To quantify the agreement between the wavelet reconstruction for a network of detectors, we compute the network match ($\match_{\rm{net}}$) as given in~\cite{Becsy:2016ofp},
\begin{equation}
\match_{\rm{net}} = \frac{ \sum_k \inp{ h^{k}_{ \mathrm{inj} } }{  h^{k}_{ \mathrm{rec} } }  }{ \left(  \sum_k\inp{ h^{k}_{ \mathrm{inj} } }{  h^{k}_{ \mathrm{inj} } } \cdot \sum_k\inp{ h^{k}_{ \mathrm{rec} } }{  h^{k}_{ \mathrm{rec} } }  \right)^{1/2} } , 
\end{equation}
where $k$ represents the $k\nth$ detector. In this paper, we consider three detectors configuration of Hanford (H1), Livingston (L1), and Virgo (V1). We consider $\aligo$ design sensitivity for H1 and L1 detectors~\cite{LIGOScientific:2014pky, Harry:2010zz, aLIGO_ZDHP}, and advanced Virgo design sensitivity for V1~\cite{Accadia_2012, VIRGO:2014yos}. Please note that we demonstrate the injection analysis in this section assuming the single detector with the $\aligo$ design sensitivity.

A whiten Gaussian noise is considered to be distributed normally over time-scale domain. This implies that the noise energy is equally distributed over time-scale domain. The amount of noise retains in the reconstructed signal is determined by the spectral loss parameter $R_\ast$. A higher value of $R_\ast$ can further reduce the inseparable noise in the reconstructed signal. At the same time, the essential wavelets would not be able to represent the complete signal characteristics. It implies that the spectral loss parameter plays a role that imposes a limitation on achieving the maximum overlap between the injected and reconstructed signal. Therefore, we want to optimize this parameter. For a given source parameter and injected SNR, we perform the injection analysis with a set of $R_\ast$ values. The noise averaged match for a given injected signal $h_{\rm{inj}}( \vec{\lambda}, \rho_{\rm{inj} } )$ and $R_\ast$ can be defined as
\begin{equation}
\label{eq:SpectralLoss}
\langle \match( R_\ast ) \rangle_{ \rm{noise} } = \frac{1}{N}\sum_{i=1}^N \match (\hinj, \hrec [n_i] ), 
\end{equation}
where, $\hrec [n_i] $ represents the reconstructed signal from the data of $i\nth$ noise realization $n_i$. 
To determine the essential wavelets for a given value of $R_\ast$, we compute the wavelet coefficients of whitened waveform and plug in to Eq.~\eqref{eq:ImportantWavelets}.

Fig.~\ref{fig:spec_loss} shows the noise averaged match as a function of $R_\ast$ for a set of injected SNR values and the number of essential wavelets as a function of $R_\ast$. If we fix the spectral loss parameter and increase the injected SNR, the signal contribution to each essential wavelet increases. At the same time, the average noise contribution remains the same since the number of essential wavelets and their properties does not change. This leads to an improvement in the reconstruction accuracy. A fixed SNR curve in the Fig.~\ref{fig:spec_loss} indicates that a smaller value of $R_\ast$ allows many nonessential wavelets in the analysis. The signal contribution to those wavelets is trivial, but the inseparable noise in the reconstructed signal increases. On the other hand, a higher value of $R_\ast$ discards many moderate essential wavelets. Thus, the choice of $R_\ast$ is a tradeoff between the signal and noise contributions to the wavelets. To find an optimum value R, we maximize the quantity $\langle \match(  R_\ast ) \rangle_{\rm{noise}} $ over $R_\ast$.


\begin{figure}
\centering
  \includegraphics[width=0.48\textwidth]{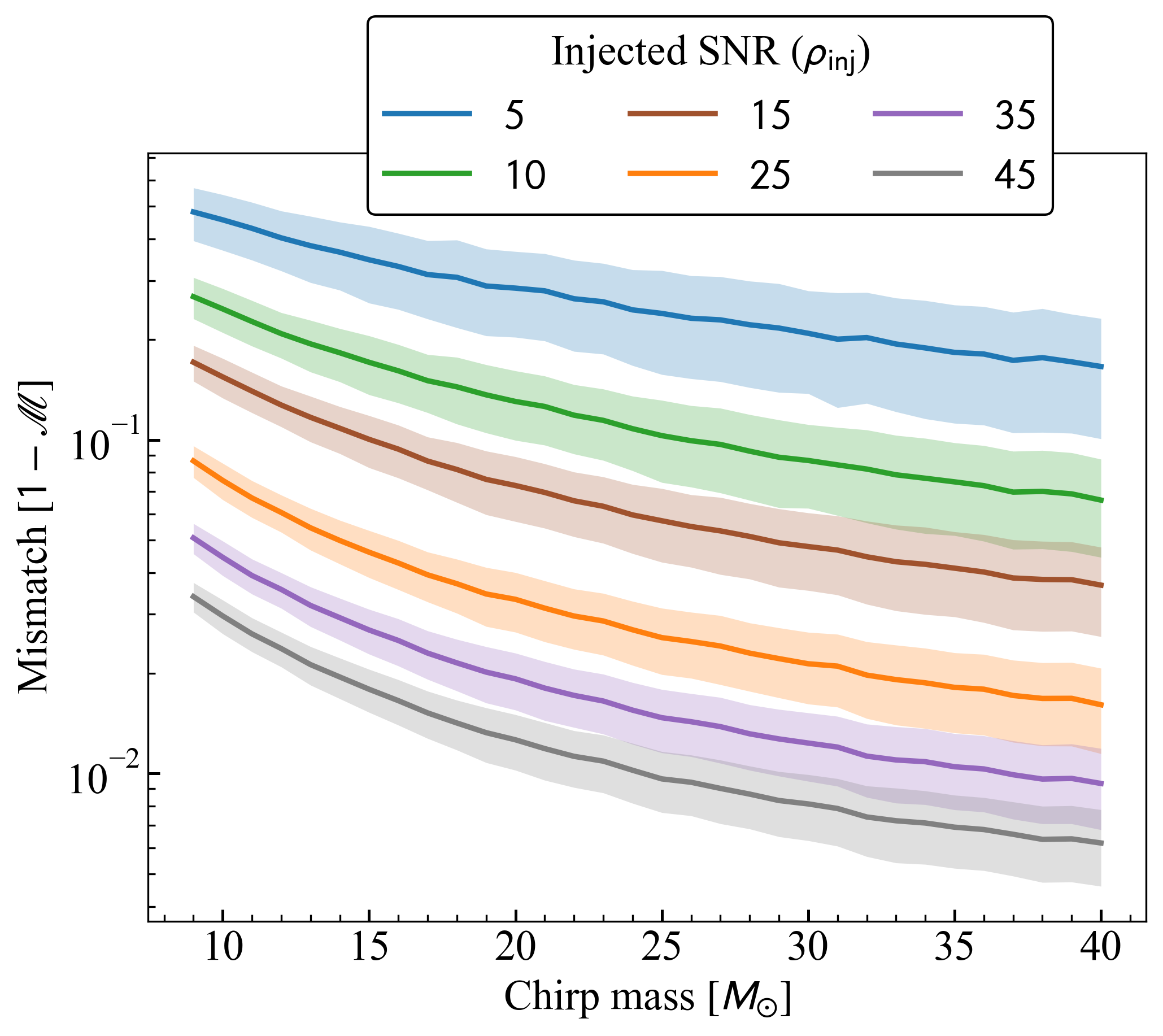}
  \caption{ The figures shows the mismatch between the injected gravitational wave signal and reconstructed signal, where the filter wavelets were constructed using the log-uniform scale as defined in Eq.~\eqref{eq:log-uniformSacle}. For producing the injections, we have assumed the nonspinning equal mass binary black holes with a wide range of chirp mass as labelled in $x$-axis. The legend of the figures stand for the optimal SNR of the injected signal. We have injected an identical signal into many noise realization to produce the distribution of mismatch for each chirp mass. We have assumed the $\aligo$ noise curve with a fixed lower cutoff frequency of $20\Hz$. The waveforms are generated using the $\seobnr$ model.} 
  \label{fig:match_comparison}
\end{figure}

Fig.~\ref{fig:match_comparison} demonstrates the \textit{mismatch} ($1 - \match $) in reconstruction for nonspinning binary systems with equal component masses. 
The solid curve represents the median of the mismatch distribution, and the shaded region shows the $\pm \sigma$ width of that distribution. For the case of a fixed optimal SNR, we can see that mismatch substantially decreases with an increase of chirp mass. It is intuitively expected: the number of essential wavelets for high chirp mass systems is fewer than low chirp mass systems.
The waveform of a binary system with high chirp mass can be characterized using a few wavelets as the waveform is short.  In contrast, the waveform of a low chirp mass system is longer, for which one requires a large number of wavelets to represent the signal.



As the injection chirp mass increases, the number of essential wavelets decreases, and they cover a smaller area over the time-frequency plane. The signal contribution to the essential wavelet coefficients increases with the injection chirp mass for a fixed SNR case. At the same time, the noise contribution decreases due to the smaller area. That explains why the reconstruction signal for a high chirp mass system is more accurate than a low chirp mass.



\begin{figure*}[t]
\centering
  \includegraphics[width=0.98\textwidth]{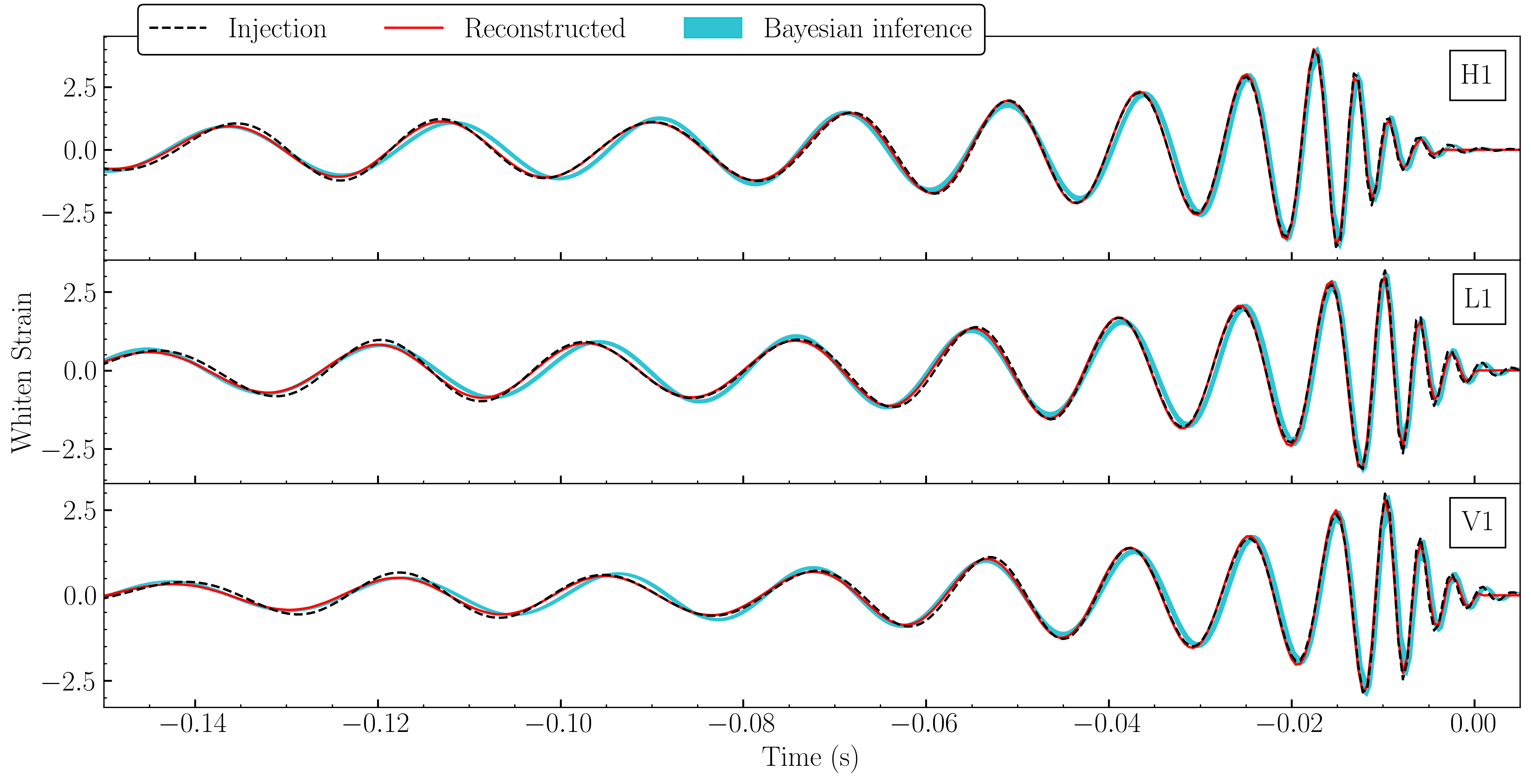}
  \caption{ An illustration of wavelet reconstruction (solid red line) and $\lalinf$ (cyan band) with  $\imrxp$  waveform model, obtained from an injection analysis using an NR eccentric waveform SXS:BBH:0323 picked from SXS catalog with a total mass of $60\msun$. The dashed black line represents the injected waveform. For this computation, we have used the three detector configuration of Advanced LIGO and Advanced Virgo. }
  \label{fig:eccnReconstruction}
\end{figure*}

\section{Identifying the deviation}
\label{sec:Deviation}
As of yet, we discussed the performance of wavelet reconstruction where injected waveforms were used for determining the essential wavelets. That results tell us the efficiency of the semi-model dependent wavelet reconstruction method when a signal in the data belongs to the search template waveform manifold. However, a signal in real data may not belong to that manifold. The deviations could arise due to: the influence of unknown binary parameters (such as eccentricity if the waveform model only considers the circular binary), missing physics in the waveform model, deviation from GR, or noise artifacts. 
We demonstrate one of such cases, an injection waveform simulation for an eccentric BBH merger. Since recent template-based analysis by LIGO-Virgo collaborations used quasi-circular waveform~\cite{GWTC2}, we consider a recently developed $\imrxp$ model to generate the template waveform~\cite{Pratten:2020ceb}. This model includes the effect of orbital precession but does not consider orbital eccentricity. 

The inclusion of eccentricity in the injected signal leads to a deviation from the search template waveform manifold. To demonstrate this, we consider a numerical-relativity simulation of eccentric IMR waveform picked from Simulating eXtreme Spacetimes (SXS) catalog~\cite{ian_hinder_2019_3326460}. This system is a nonspinning binary with mass ratio of 1.22. The numerical simulation is performed for an eccentric BBH system, where eccentricity evolves with time. The reference eccentricity is $e_{\rm{ref}} = 0.194$ as measured at a reference orbital frequency of $Mf_{0}=0.0137$. In our analysis, we scale the simulation for a total mass of $60 \, \msun$, which is suitable for ground based detectors.

We keep the system in the \textit{face-on} configuration where the inclination angle is $0^{\circ}$. As the component masses are nearly equal and zero inclination, the contribution of higher-order modes to the injected signal is negligible. This configuration assures us that the deviation enters only due to the orbital eccentricity. We inject the signal assuming a three detector configuration of Advanced LIGO and Advanced Virgo~\cite{LIGOScientific:2014pky, Harry:2010zz, aLIGO_ZDHP, Accadia_2012, VIRGO:2014yos}. We set the source's sky position and distance such that the injected SNR in H1, L1 and V1 are 25, 20, and 15, respectively. Since the average result of our wavelet reconstruction over many Gaussian noise realizations with an identical injected signal leads to the case of zero noise, we consider zero noise realization to construct the data stream.  We analyze the data using the standard Bayesian parameter estimation library LALInferenceNest~\cite{Veitch:2014wba}, a Bayesian inference nested sampling code implemented in the LIGO Algorithm Library (LALSuite)~\cite{lalsuite}. We use the python-based package $\pesummary$~\cite{PESummary} to process the data from parameter estimation analysis and generate the template waveform in the detector frame.

In order to reconstruct the signal from data stream, we follow these steps:
\begin{enumerate}
\label{enu:Reconstruction}
\item \label{en:step1}  For each posterior sample, we generate the CBC template waveform in the detector frame. Since the GR allows only two polarization states, referred to as the plus ($h_+$) and cross ($h_\times$), the time-domain response $h_\mathcal{I}(t)$ of a given detector $\mathcal{I}$ is determined by the antenna response functions ($F^+_{\mathcal{I}}$ and $F^\times_{I}$) of those polarizations~\cite{LIGOScientific:2019hgc},
\begin{equation}
\begin{split}
h_{\mathcal{I}}(t) &=  F^+_{\mathcal{I}}(\alpha, \delta, \psi, t) h_\times(t-\Delta t_{\mathcal{I}}; D_L, \iota, \vec{\lambda}) \\
 & + F^\times_{\mathcal{I}} (\alpha, \delta, \psi, t) h_+(t-\Delta t_{\mathcal{I}}; D_L, \iota, \vec{\lambda}),
\end{split} 
\end{equation}
where, $\alpha$ and $\delta$ refers to the source sky location in terms of right ascension and declination, $\psi$ is the polarization angle, $D_L$ is the luminosity distance to the source, $\iota$ is the inclination angle of the binary plane, $\vec{\lambda}$ represents the set of intrinsic parameters of the binary system, and $\Delta t_{\mathcal{I}} (\equiv  \Delta t_{\mathcal{I}} (\alpha, \delta, t) )$ is the travel time of the signal from geocenter to the detector.


\item Whiten the waveform weighted by the noise amplitude spectral density such that the norm of the whitened waveform is equal to its optimal SNR.


\item Compute the wavelet coefficients for each whiten template waveform using the CWT as shown in Eq.~\eqref{eq:cwt},, where wavelets are constructed using the log-uniform scale.


\item Determine the essential wavelets using Eq.~\eqref{eq:ImportantWavelets}, where spectral loss parameter  $R_\ast$ is obtained from Eq.~\eqref{eq:SpectralLoss}. We inject the best fit template (corresponds to maximum likelihood sample) waveform in many simulated noise realizations and estimate the value of  $R_\ast$ for each detector’s data. We have found the value of  $R_\ast$ for H1, L1, and V1 is 2.5\%, 4\%, and 5\%, respectively. Please note that we estimate the spectral loss parameter for the maximum likelihood sample and use that value for all the posterior samples.



\item Use those essential wavelets to reconstruct the signal from detector strain.


\end{enumerate}  

\begin{table}[htbp!]
\centering
\begin{tabular}{l | c c c c c   }
\toprule[1pt]
\toprule[1pt]
Detector & \ $\rho_{\rm{inj}}$  \ & \   $ \widetilde{ \match}_{\rm{bif}}^{\rm{inj}}$ \ & \ $\widetilde{ \match}^{\rm{rec}}_{\rm{bif}}  $ \ & \ $ \widetilde{\match}_{\rm{rec}}^{\rm{inj}} $ \ & \ $\widetilde{\rho}_{\rm{res}}$    \\
\midrule[1pt]
Hanford (H1) & 25 & 0.954  & 0.968 & 0.984 & 6.2  \\
Livingston (L1) & 20 &   0.95 & 0.959 & 0.978 & 5.5 \\
Virgo (V1)       & 15 &   0.959 & 0.959 & 0.978 & 4.2 \\
Network         & 35.6 & 0.949 & 0.958 & 0.981 & -- \\
\bottomrule[1pt]
\bottomrule[1pt]
\end{tabular}
\caption{ Comparison between the $\lalinf$ template waveform and log-uniform wavelet reconstruction for an eccentric numerical relativity waveform SXS:BBH:0323 picked from SXS catalogue with a total mass of $60 \msun$. The numerical simulation is performed for a nonspinning, nearly equal-mass eccentric BBH system, where the measured reference eccentricity is $e_{\rm{ref}} = 0.194$ at a reference orbital frequency of $Mf_{0}=0.0137$. The quantity $\widetilde{ \match}$ denotes the median of the match values.}
\label{tab:matchComparison}
\end{table}

The above-described procedure is used to obtain a reconstructed signal for each posterior sample. These wavelet-based reconstructed signals are very similar, and their 90\% interval is very thin and looks like a line. Therefore, we illustrate the median of reconstructed signals at every time index.~\footnote{Note that the median of the reconstructed signals is used \emph{only} for the illustration in time-domain.} 
Fig.~\ref{fig:eccnReconstruction} illustrates the results from wavelet reconstruction, 90\% credible region of $\lalinf$ template waveform, and injected waveform. We can see the amplitude and phase of the reconstructed waveform are approximately consistent with the injected waveform. On the contrary, the $\lalinf$ waveform is out of phase over a bit of the region, which indicates a significant deviation from the search template waveform manifold. We also compute three different matches using the reconstructed signal ($\hrec$), $\lalinf$ template waveform ($\hlif$) obtained from posterior samples as described in step~\ref{en:step1}, and injected waveform ($\hinj$): $\match(\hinj, \hlif)$, $\match(\hlif, \hrec)$, and $\match(\hinj, \hrec)$. 
Table~\ref{tab:matchComparison} summarizes the match comparison.  It signifies the wavelet-based reconstructed waveform is more faithful than the $\lalinf$. 

\begin{figure*}
\centering
  \includegraphics[width=0.98\textwidth]{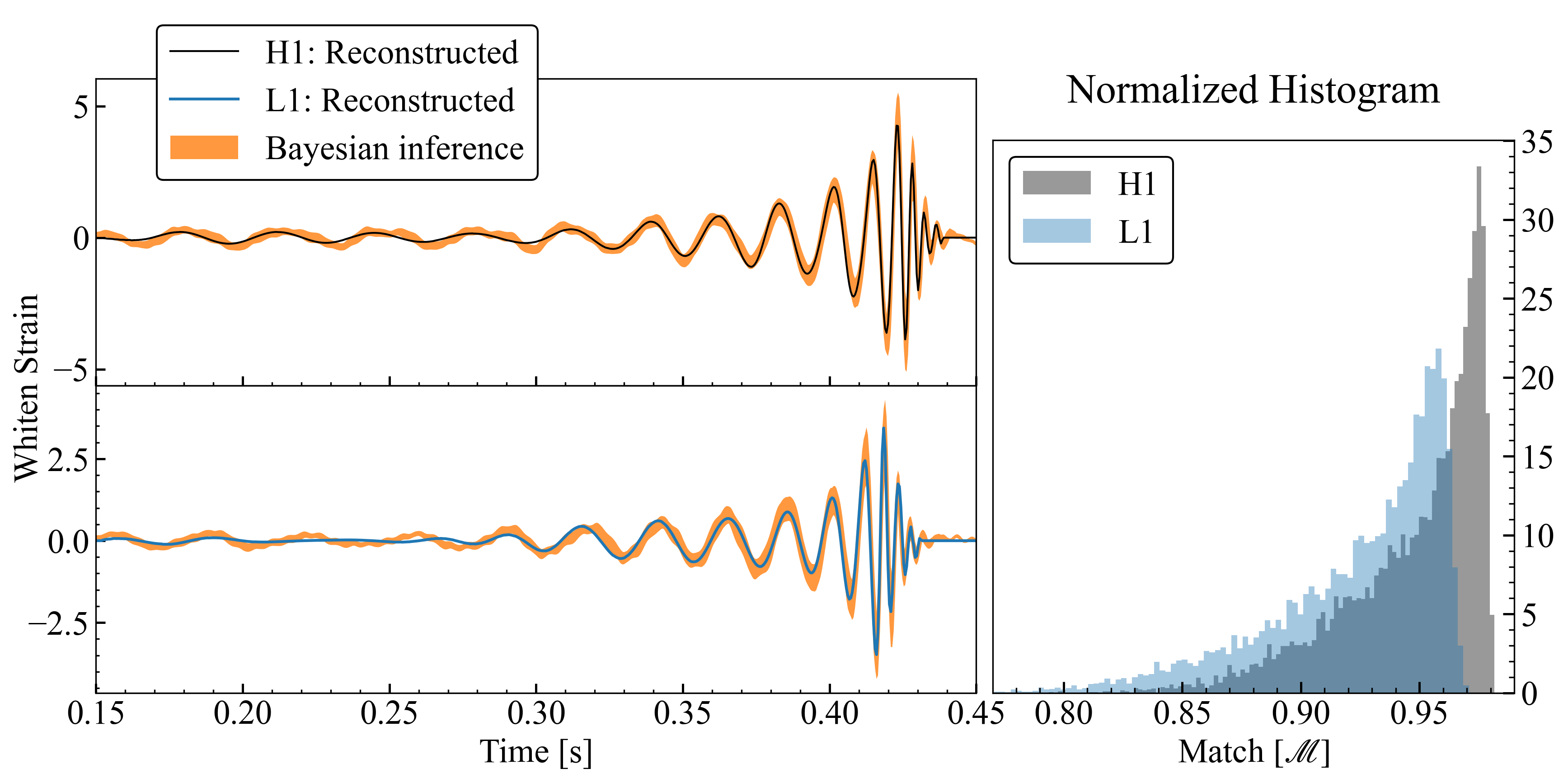}
  \caption{ Log-uniform scale wavelet reconstruction results of GW150914 event, obtained using the on-source data of H1 and L1 detectors, where the times (in seconds) are shown relative to a reference time 1126259462.0. The solid black line in the top-left panel shows the median reconstructed signal obtained from H1 data and the solid blue line in the bottom-left panel for L1 data. The orange band in these panels is produced using the CBC template waveforms from parameter estimation samples. We demonstrate the reconstructed signals and CBC template waveforms in units of the standard deviation of the noise, which implies the norm is SNR. The right panel shows the histogram of the match between every posterior waveform and the corresponding wavelet-based reconstructed signal. }
  \label{fig:gw150914Reconstruction}
\end{figure*}

However, we can not compute the $\match(\hrec, \hinj)$ and $\match(\hlif, \hinj)$ for an actual event case as it is infeasible to know the true signal in data. We propose to compute the residual SNR ($\rho_{\rm{res}}$) obtained by subtracting each template waveform of $\lalinf$ posterior samples from the corresponding wavelet- based reconstructed waveform. LVC commonly uses this procedure in the test of GR with BBH events~\cite{TGR-GW150914, TGR-GWTC1, TGR-GWTC2, 2021arXiv211206861T}. We report the median of residual SNR in the last column of Table~\ref{tab:matchComparison}. 



\section{Analysis of events in GWTC-1}
\label{sec:GWTC1Ananlysis}

We apply the proposed method to each binary black hole event in the first gravitational-wave transient catalog GWTC-1~\cite{GWTC1}  to reconstruct the signals from individual detectors. In this analysis, we use the on-source data from the Gravitational Wave Open Science Center~\cite{GWOSC:catalog, LIGOScientific:2019lzm}, released for GWTC-1~\cite{GWTC1}. For determining the essential wavelets, we use the posterior samples of source properties obtained from $\bilby$'s~\cite{Ashton:2018jfp, Romero-Shaw:2020owr} reanalysis of GWTC-1. Parameter estimation analysis was performed using the $\imrppv$~\citep{Schmidt:2012rh, Hannam:2013oca} waveform model, and PSD was estimated using the $\bayesline$ algorithm~\cite{Littenberg:2014oda}.

In order to reconstruct the signal from data using the posterior samples, we follow the steps described in Section~\ref{sec:Deviation}. Fig.~\ref{fig:gw150914Reconstruction} shows the results of GW150914~\cite{gw150914}: CBC template waveform from posterior sample and wavelet reconstruction. 
The agreement for H1 data is better than L1 data as the reconstruction accuracy increases with SNR. For H1 data, the most probable value (mode) of the match values is 0.975, and the maximum is 0.982. We also compute the network match values and its mode 0.962. It implies an excellent agreement between the GR template waveform and the observed data. We have seen that the reconstructed signal does remain almost identical even when the essential wavelets are selected using different posterior samples. Therefore, we consider the case of the maximum likelihood sample only to illustrate the signal is time-domain. Note that match values are computed between every posterior waveform and the corresponding wavelet-based reconstructed signal; we call this \emph{on-source} match.


Our reconstruction method trades with sensitivity for identifying the essential wavelets and removing the noise from on-source data. Consequently, the technique cannot discern the early inspiral or late ringdown part of the CBC waveforms where the signals are weaker. In the time-frequency domain, the early inspiral part spreads over time, whereas the late-ringdown part of the signal over frequency direction. In Fig.~\ref{fig:gw150914Reconstruction}, it is visible that the early inspiral part of the L1 signal (before 0.25 seconds) fades out. However, the H1 signal is still present and consistent with the template waveform because of the higher SNR in H1. At the same time, the late ringdown part of both the signals is disappeared. In a time-frequency frame, the ringdown part of a signal spreads out over frequency despite having a fixed frequency because of its exponential decay term, leading to a Lorentzian spread along the frequency direction.

\begin{figure}
\centering
  \includegraphics[width=0.48\textwidth]{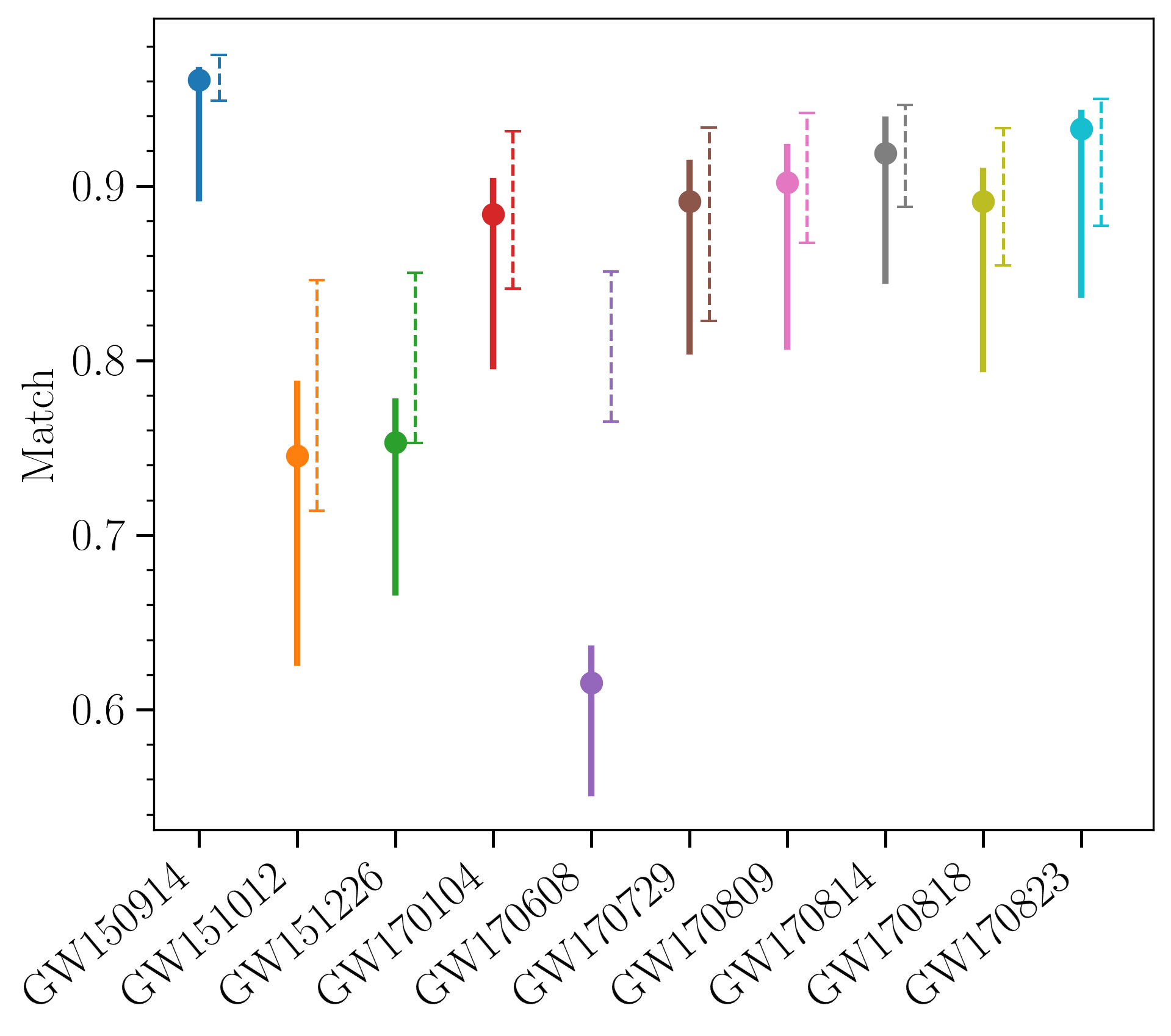}
  \caption{ Network match between the CBC template waveform by Bayesian parameter estimation and wavelet-based reconstructed waveform. The vertical solid line represents the 90\% credible interval of the on-source match. The dot over each line is the mode of the on-source match distribution. The vertical dashed line indicates the 90\% interval of the expected match under the assumption of Gaussian noise, which is similar to Fig.~\ref{fig:match_comparison}.}
  \label{fig:gwtc1_match}
\end{figure}

\begin{table}
\centering
\begin{tabular}{c | c c c c c c c c }
\toprule[1pt]
\toprule[1pt]
Event \ &  $\widetilde{\mathcal{M}}_c$ & \ $\widetilde{ \rho}_{\rm{bif}}^{\: \rm{H1}}  $ & \ $\widetilde{ \rho}_{\rm{bif}}^{\: \rm{L1}}  $ & \ $\widetilde{ \rho}_{\rm{bif}}^{\: \rm{V1}}  $ \ & \ $ \widetilde{ \match}_{\rm{H1}}  $ & \ $ \widetilde{ \match}_{\rm{L1}} $ & \ $ \widetilde{ \match}_{\rm{V1}} $ & \ $ \widetilde{ \match}_{\rm{net}} $    \\
\midrule[1pt]
  GW150914  & 31.0 & 20.6 & 14.3 & --   & 0.96 & 0.94 & -- & 0.95\\
  GW151012 & 18.3 & 6.5 & 5.9    & -- & 0.73 & 0.74 & -- & 0.73\\ 
  GW151226  & 9.7   & 9.8 & 6.9 & --   & 0.80 & 0.66 & --  & 0.74 \\
  GW170104  & 25.7 & 9.5 & 10.0 & --   & 0.86 & 0.89 & -- & 0.87 \\
  GW170608  & 8.5   & 12.1 & 9.2 & --      & 0.78 & 0.49 & -- & 0.60   \\ 
  GW170729  & 51.5 & 6.0  & 8.3 & 1.7   & 0.89 & 0.91 & -0.21 & 0.85    \\
  GW170809  & 29.7 & 6.0 & 10.8 & 1.1  & 0.86 & 0.92 & 0.02 & 0.89 \\
  GW170814  & 27.0 & 9.3 & 14.2 & 3.8  & 0.88 & 0.93 & 0.74 & 0.91 \\
  GW170818  & 32.1 & 4.7 & 9.8 & 4.3 & 0.80 & 0.90 & 0.82 & 0.87   \\
  GW170823  & 38.9 & 7.1 & 9.5 & --   & 0.88 & 0.93 & -- & 0.91 \\
\bottomrule[1pt]
\bottomrule[1pt]
\end{tabular}
\caption{ List of match values of GWTC-1 events obtained by computing the match between the reconstructed waveform and Bayesian inference template waveform. We report the chirp mass and the SNR to indicate the efficiency of the reconstruction as demonstrated in section~\ref{subsec:gwsimnoise}.  We report the median (denoted by tilde) value of the distribution. Dashes (–) correspond to detector not included in the analysis.  }
\label{tab:matchComparisonGWTC-1}
\end{table}


Further, we perform the reconstruction analysis on the remaining events of GWTC-1~\cite{LIGOScientific:2016sjg, LIGOScientific:2016dsl, LIGOScientific:2017bnn, LIGOScientific:2017vox, LIGOScientific:2017ycc} and report the results in Table~\ref{tab:matchComparisonGWTC-1}. In order to understand the reconstruction efficiency depending on the SNR and chirp mass, we also reported them in the same Table. These values are obtained from the posterior samples by $\bilby$~\cite{Romero-Shaw:2020owr}. The solid vertical line in Fig.~\ref{fig:gwtc1_match} shows the 90\% interval of the on-source network match values, and the solid circle marks their mode value. We found the best agreement with GR for the GW150914 event, the minimum for GW170608, albeit the observed SNR from the latter event was higher than the nominal threshold in both the detectors. The match value depends not only on SNR but also on the time-frequency area covered by the essential wavelets of the signal. In section~\ref{subsec:gwsimnoise}, we have seen that the match value increases with injection chirp mass while the injection SNR is kept at a fixed value. GW170608 event has the lowest chirp mass in the catalog, for which its essential wavelets occupy the largest area over the time-frequency plane.

The vertical dashed line in Fig.~\ref{fig:gwtc1_match} shows the 90\% interval of expected match that is determined by injecting the maximum likelihood $\bilby$ waveform ($\hlif^\ast$) in many Gaussian noise realizations, similar study is shown in Fig.~\ref{fig:match_comparison}. We have reconstructed the signal from each realization and computed the match with $\hlif^\ast$. We see that there is significant overlap between the distribution of expected match and on-source match, except for GW151226 and GW170608. In particular, the distributions are far from each other for GW170608, for which a further study is worthy. We will focus on it in future work. 
\begin{figure}
\centering
  \includegraphics[width=0.48\textwidth]{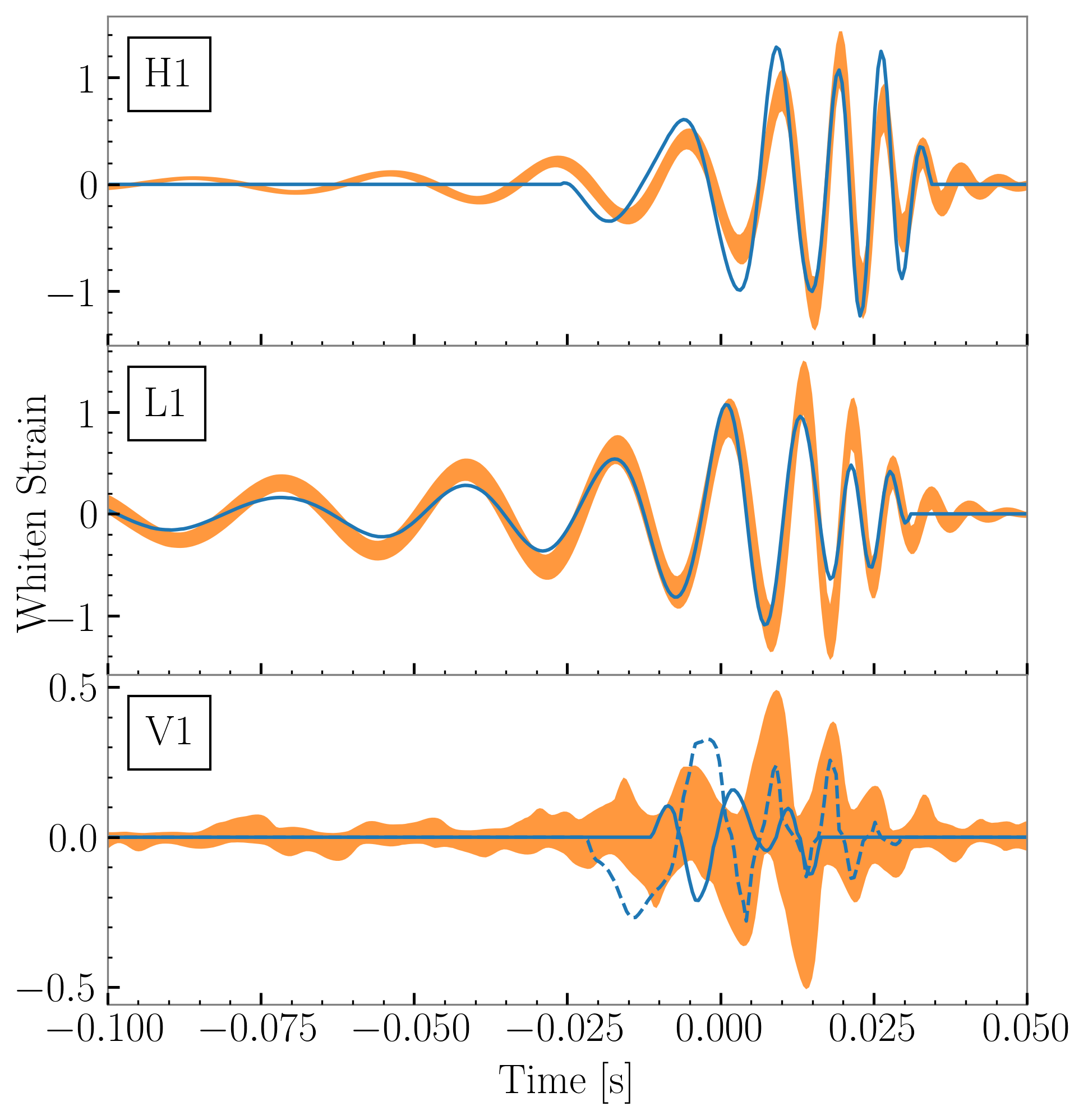}
  \caption{Signal reconstruction results of GW170729 with a reference time 1185389807.3. We plot the 90\% credible interval of the CBC template waveform (orange band) and median reconstructed signal (blue). The blue dashed line in the bottom panel shows the projected signal of H1 and L1 over V1, where the sky location and polarization angle are taken from parameter estimation samples. }
  \label{fig:gw170729_rec}
\end{figure}

For GW170729, the match V1 data is negative, which is unexpected. The reconstructed signal and template waveform are in nearly antiphase for most of the posterior samples as shown in Fig.~\ref{fig:gw170729_rec}, leading to a negative match value. It could be due to the noise artifacts. The antiphase could occur due to poor constraint of event time in the detector frame. The total width of event time distribution for H1, L1, and V1 are 19, 18, and 63 ms, respectively. The spread for V1 data is significantly larger than the other two. It is impossible to draw an appropriate conclusion about whether the deviation presents or not when the SNR is very low. Thereby, we also compute the network match excluding the contribution of V1. The median network match for two detectors is 0.88, and the mode is 0.89. An extensive study on agreement between the CBC template waveform and the $\bayeswave$ reconstruction also found the similar results~\cite{Chatziioannou:2019dsz}, where the contribution of V1 data was not considered for this study. We excluded the contribution of V1 for GW170729 to produce Fig.~\ref{fig:gwtc1_match}. As a further investigation, we project the H1 and L1 signal over V1 based on the sky location and polarization angle of parameter estimation samples. We follow these steps to obtain the projected signal: (a) perform the inverse transformation of whiten procedure over H1 and L1 reconstructed signals and apply time-shift based on the sky location to obtain these signals in geocentric coordinate,  (b) obtain the two GW polarizations based on the sky location and polarization angle, and (c) project on the frame of V1 detector as shown in the bottom panel (blue dashed) of Fig.~\ref{fig:gw170729_rec}. We can see that the projected signal and CBC template waveform are in phase, and the median match is 0.64. It implies that the noise probably plays a role in V1 reconstructed signal being antiphase.




\section{Conclusion}
\label{sec:Consclusion}
This paper presents a wavelet-based semi-model-dependent method for reconstructing the gravitational wave signals produced from compact binary mergers. We have employed the framework of continuous wavelet transformation, where Morlet wavelets represent the signals. The semi-model-dependent approach determines the essential wavelets using the posterior samples from parameter estimation to reconstruct the signals from the data.

In general, the wavelets are constructed using an output of octave scale, which provides the tightest set of wavelets. Such wavelets yield a nearly orthogonal wavelet basis. In this paper, we have proposed a log-uniform scale for constructing the wavelets. Such wavelets are highly redundant, i.e., non-orthogonal wavelets. However, this new scale is more efficient for representing the linear chirp signals at high frequencies than the octave scale. As the wavelets are regarded as time localized bandpass filters, a reconstructed signal always contains a nominal amount of noise that passes through the essential wavelets. The amount of noise depends on the area covered by the essential wavelets over the time-frequency plane. We have shown that the essential wavelets with the log-uniform scale occupy a smaller space than the octave scale, which enables us to reconstruct the weak signals better. 

We have conducted injection analysis to evaluate the reconstruction efficiency for the signals from binary black hole mergers by computing the match between injected waveform and the reconstructed signal. The reconstruction accuracy increases with the SNR such that the mismatch is approximately proportional to SNR squared, $1 - \match \gtrsim 1/\mathrm{SNR}^2$. Also, the reconstruction accuracy strongly depends on chirp mass. We have seen that the mismatch at a fixed SNR decreases as chirp mass increases. 

We have demonstrated the ability to detect the deviation where the injected waveform is outside the region of space enclosed by the search template waveform manifold. We have performed the parameter estimation analysis using $\lalinf$ by injecting an electric BBH waveform, where the template waveforms are generated by a quasi-circular waveform model $\imrxp$. After that, we studied the match between the injected waveform and the reconstructed waveform. As reported in Table~\ref{tab:matchComparison}, the $\lalinf$ waveform agrees less with the injected waveform than the wavelet reconstruction. Also, the $\lalinf$ waveform is out of phase over a bit of the region. In comparison, the amplitude and phase of the reconstructed waveform are nearly consistent with the injected waveform.


We have applied this wavelet-based reconstruction analysis to each binary black hole merger event in GWTC-1. We have seen a satisfactory agreement between the reconstructed signal and the estimated theoretical waveform. As the gravitational wave catalog grows, we expect the wavelet-based semi-model-dependent reconstruction method to provide a more precise view of the agreement between the observed data and the waveform model. 

There are many avenues for future work and extensions of this method: combining the octave and log-uniform scales to have an optimal method for reconstructing the signals from compact binaries; applying this new method for reconstructing the signal from binary neutron star mergers; developing appropriate statistics for detecting the deviations and investigating the match study when instrumental glitch presents in on-source data; applying this method on GWTC-2 and GWTC-3~\cite{GWTC2, 2021arXiv211103606T}.

\begin{acknowledgments}

I gratefully acknowledge  Ayatri Singha, Sudarshan Ghonge, M.K. Haris, Amit Reza, Khun Sang Phukon, Chinmay Kalaghatgi, Bhooshan Gadre, Melissa Lopez Portilla, Chris Van Den Broeck for helpful comments and suggestions. S.R. was supported by the research program of the Netherlands Organization for Scientific Research (NWO). The material of this paper is based upon work supported by NSF's LIGO Laboratory, which is a major facility fully funded by the National Science Foundation (NSF). I gratefully acknowledge computational resources provided by the LIGO Laboratory and supported by the NSF Grants No.~PHY-0757058 and No.~PHY-0823459. This research has made use of data, software and/or web tools obtained from the Gravitational Wave Open Science Center, a service of LIGO Laboratory~\cite{GWOSC:catalog}, the LIGO Scientific Collaboration and the Virgo Collaboration. 

\end{acknowledgments}

\bibliography{reference}{}

\end{document}